\documentstyle[12pt,epsfig]{article}
\def\lbldef#1#2{\expandafter\gdef\csname #1\endcsname {#2}}

\def\href#1#2{#2}  

\begin{document}
\baselineskip=15.5pt
\pagestyle{plain}
\setcounter{page}{1}

\begin{titlepage}

\begin{flushright}
CERN-TH-194\\
hep-th/9906232
\end{flushright}
\vspace{10 mm}

\begin{center}
{\Large Probing partially localized supergravity 
background of fundamental string ending on D$p$-brane}

\vspace{5mm}

\end{center}

\vspace{5 mm}

\begin{center}
{\large Donam Youm\footnote{Donam.Youm@cern.ch}}

\vspace{3mm}

Theory Division, CERN, CH-1211, Geneva 23, Switzerland

\end{center}

\vspace{1cm}

\begin{center}
{\large Abstract}
\end{center}

\noindent

We study the dynamics of the probe fundamental string in the field background 
of the partially localized supergravity solution for the fundamental string 
ending on D$p$-brane.  We separately analyze the probe dynamics for its 
motion along the worldvolume direction and the transverse direction of the 
source D$p$-brane.  We compare the dynamics of the probe along the D$p$-brane 
worldvolume direction to the BIon dynamics.

\vspace{1cm}
\begin{flushleft}
CERN-TH-194\\
June, 1999
\end{flushleft}
\end{titlepage}
\newpage

\section{Introduction}

The source and probe method has been useful for studying the bound states 
of branes.  In this method, it is assumed that the source brane is much 
heavier than the probe brane.  Thus, although the probe is under the 
influence of the fields produced by the source, the probe has no influence 
on the source field configuration.  So, the source and probe system is 
described by the Dirac-Born-Infeld (DBI) or the Nambu-Goto (NB) action of 
the probe brane in the `static' field background of the source brane.  Such 
method has been successful in reproducing the brane intersection rules 
\cite{tse1} and in studying some dynamics or statistical mechanics of branes, 
e.g. Refs. \cite{dps,tse2,tse3,tse4,tse5}.  

The source and probe method is useful especially when one wants to study 
intersecting brane configuration, since completely localized supergravity 
solution for intersecting branes are not yet available.  Note, for 
delocalized supergravity solutions, a constituent brane is not localized on 
the worldvolume of the other constituent.  So, delocalized solutions are not 
useful for studying, for example, the dynamics of a brane constituent within 
the worldvolume of another brane constituent.  For such study, one lets one 
constituent to be the probe and another constituent to be the `static' 
background in which the probe moves.  However, this method cannot be applied 
if one also wants to study the interaction among branes of the same type 
while these branes move in the background of brane of another type.  It is 
the purpose of this paper to study such case. 

In Refs. \cite{tse6,youm,fs,loe}, various types of (partially) localized 
supergravity solutions for intersecting branes in the core region of one 
constituent brane are constructed.  For such solutions, one constituent 
is localized on the worldvolume of the other constituent (while the latter 
brane is delocalized on the former brane), provided that some of the overall 
transverse directions are delocalized in some cases.  In this paper, 
we study the dynamics of the (localized) former type of brane (brane 1), 
which not only interacts with the (delocalized) other type of brane (brane 
2) but also interacts with another brane 1.  One can describe such dynamics 
with the DBI or the NG action of the brane 1 in the background field 
configuration of partially localized intersecting brane 1 and brane 2, 
where brane 1 is localized on brane 2.  

One can apply this method to study the dynamics for any type of intersecting 
brane configurations by using the partially localized intersecting brane 
solutions and applying the similar procedure as the work of this paper.  
But we will restrict our study to the case of the fundamental strings ending 
on D$p$-brane, because for this case the dynamics of the corresponding 
worldvolume solitons, i.e. the BIons \cite{cm,gib}, is relatively 
well-understood, e.g. Refs. \cite{gp1,blm,gp2}.  As we will see in the 
following section, we however find disagreement in dynamics of the probe 
fundamental string with the dynamics of BIons.  This seems to be due to the 
fact that the supergravity solutions for intersecting branes used in this 
paper is not fully localized ones.  Namely, for the partially localized 
intersecting brane solutions used in this paper, the location of D$p$-brane 
along the longitudinal direction of the fundamental string is not specified, 
whereas the scalar field of the BIon solution specifies such location.  Also, 
it might be due to the difference in approximations used in the calculation of 
brane dynamics in the source-probe method and the worldvolume soliton method.  
However, the comparison of supergravity and worldvolume aspects of brane 
dynamics discussed in this paper may turn out to be useful in other relevant 
studies and when fully localized solutions are available.   

The paper is organized as follows.   In section 2, we summarize the 
partially localized supergravity solution for the fundamental string ending 
on D$p$-brane.  In section 3, we study the dynamics of the probe fundamental 
string in this supergravity background, closely following the previous 
works \cite{dps,tse4} on the dynamics of the probe branes.  We consider the 
cases where probe fundamental string moves along the longitudinal 
direction and the transverse direction of the source D$p$-brane, separately.

\section{Partially Localized Supergravity Solution for the Fundamental 
String Ending on D$p$-brane}

In this section, we summarize the partially localized supergravity solution 
for the fundamental string ending on D$p$-brane
\footnote{Of course, this supergravity solution does not strictly correspond 
to the fundamental string `ending' on D$p$-brane, but rather corresponds to 
the fundamental string `piercing' through D$p$-brane.  But at this moment, 
this supergravity solution is the closest that we have.}
constructed in Ref. \cite{youm}.  The string-frame effective supergravity 
action for such configuration is given by
\begin{equation}
S={1\over{16\pi G_{10}}}\int d^{10}x\sqrt{-G}[e^{-2\phi}({\cal R}+
4\partial_M\phi\partial^M\phi-{1\over{12}}|H_3|^3)-{1\over{2(p+2)!}}
|F_{p+2}|^2],
\label{sglag}
\end{equation}
where $G$ and ${\cal R}$ are respectively the determinant and the 
Ricci scalar of the spacetime metric $G_{MN}$ ($M,N=0,1,...,9$) in the 
string frame, $\phi$ is the dilaton, $H_3$ is the field strength for the 
the 2-form potential $B_{MN}$ in the NS-NS sector, and $F_{p+2}$ is the field 
strength for the $(p+1)$-form potential $A_{M_1\cdots M_{p+1}}$ in the R-R 
sector.  For the $p=2$ case, there is an additional Chern-Simon term 
$\sim \epsilon^{M_1\cdots M_{10}}B_{M_1M_2}\partial_{M_3}A_{M_4M_5M_6}
\partial_{M_7}A_{M_8M_9M_{10}}$ in the action.  

The supergravity solution has the following form
\footnote{Similar class of solutions was first constructed in Ref. 
\cite{tse6}}:
\begin{eqnarray}
G_{MN}dx^Mdx^N&=&-H^{-1}_FH^{-{1\over 2}}_pdt^2+H^{-{1\over 2}}_p(dx^2_1
+\cdots+dx^2_p)
\cr
& &H^{-1}_FH^{1\over 2}_pdy^2+H^{1\over 2}_p(dz^2_1+\cdots+dz^2_{8-p}),
\cr
e^{\phi}&=&H^{-{1\over 2}}_FH^{{3-p}\over 4}_p,\ \ \ \ 
B_{ty}=-H^{-1}_F,\ \ \ \  A_{tx_1\cdots x_p}=-H^{-1}_p,
\label{f1dpsgsol}
\end{eqnarray}
where the harmonic functions for the fundamental string and the D$p$-brane 
in the near horizon region ($|\vec{z}-\vec{z}_0|\approx 0$) of the D$p$-brane 
are respectively given by
\begin{equation}
H_F=1+\sum_i{{Q_i}\over{[|\vec{x}-\vec{x}_{0\,i}|^2+{{4Q_p}\over{(p-4)^2}}
|\vec{z}-\vec{z}_0|^{p-4}]^{{(p-3)^2+3}\over{2(p-4)}}}},\ \ \ \  
H_p={{Q_p}\over{|\vec{z}-\vec{z}_0|^{6-p}}}.
\label{f1dpharms}
\end{equation}
Note, these harmonic functions describe the localized fundamental strings 
on the D$p$-brane only for the $p=5$ case.  (When $p=6$, the harmonic 
function $H_p$ is logarithmic and therefore this supergravity solution is 
not valid.  For the $p=7$ case, the spacetime is not asymptotically flat.)  
For the $p<5$ case, one has to delocalize $5-p$ of the overall transverse 
directions in order to localize the fundamental strings on the D$p$-brane.  
Harmonic functions for this case are given by
\begin{equation}
H_F=1+\sum_i{{Q_i}\over{[|\vec{x}-\vec{x}_{0\,i}|^2+4Q_p|\vec{z}-
\vec{z}_0|]^3}},\ \ \ \ 
H_p={{Q_p}\over{|\vec{z}-\vec{z}_0|}}.
\label{pless5harms}
\end{equation}
Note, in the expressions for harmonic functions in Eq. (\ref{pless5harms}), 
$\vec{z}$ is the part of overall transverse coordinates where the branes 
are localized.  Namely, $\vec{z}$ in Eq. (\ref{pless5harms}) is 
three-dimensional.

In this paper, we consider the case in which all the fundamental strings 
coincide at the origin of the D$p$-brane worldvolume space (i.e. 
$\vec{x}_{0\,i}=\vec{0}$, for all $i$) and the fundamental strings and 
the D$p$-brane meet at the origin of the overall transverse space 
(i.e. $\vec{z}_0=\vec{0}$).  The harmonic functions (\ref{f1dpharms}) 
in this case take the following forms:
\begin{equation}
H_F=\left\{\matrix{1+{{Q_F}\over{[x^2+4Q_pz]^3}},\ \ \ p<5 \cr 
1+{{Q_F}\over{[x^2+4Q_5z]^{7\over 2}}},\ \ \ p=5}\right.,\ \ \ \ 
H_p={{Q_p}\over z},
\label{coinharms}
\end{equation}
where $x\equiv |\vec{x}|$ and $z\equiv |\vec{z}|$.

\section{Dynamics of the Probe Fundamental String}

In this section, we study the dynamics of the probe fundamental 
string that moves in the background of the source fundamental string 
ending on the source D$p$-brane with the field configuration given by 
Eq. (\ref{f1dpsgsol}).  We will assume that ($i$) the source brane is 
much heavier than the probe brane, i.e. there are large numbers of 
coinciding source fundamental strings and source D$p$-branes, and ($ii$) 
the velocity of the probe fundamental string is very small and changes 
very slowly.  Based on the first assumption, we neglect the backreaction 
on the source due to the moving probe.  The second assumption implies that 
the radiation will be negligible, allowing quasistatic evolution of the 
system which is described by the geodesic motion in the moduli space.  

The action for the probe fundamental string with the tension $T_f$ moving 
in the curved background is given by the following Nambu-Goto action:
\begin{equation}
S=\int d\tau d\sigma L=-T_f\int d\tau d\sigma \left[\sqrt{-\det\hat{G}_{ab}}
+{1\over{2!}}\epsilon^{ab}\hat{B}_{ab}\right],
\label{f1probact}
\end{equation}
where $\hat{G}_{ab}$ and $\hat{B}_{ab}$ ($a,b=\tau,\sigma$) are respectively 
the pull-backs of the spacetime metric $G_{MN}$ and the NS-NS 2-form potential 
$B_{MN}$ to the worldvolume of the fundamental string, namely
\begin{equation}
\hat{G}_{ab}\equiv G_{MN}\partial_aX^M\partial_bX^N,\ \ \ \ 
\hat{B}_{ab}\equiv B_{MN}\partial_aX^M\partial_bX^N.
\label{pullbacks}
\end{equation}
In the static gauge ($X^0=\tau$ and $X^1=\sigma$ with $X^1$ being the 
longitudinal coordinate of the fundamental string), the pull-backs take 
the following forms:
\begin{equation}
\hat{G}_{ab}=G_{ab}+G_{ij}\partial_aX^i\partial_bX^j,\ \ \ \ 
\hat{B}_{ab}=B_{ab}+B_{aj}\partial_bX^j+B_{ib}\partial_aX^i,
\label{statpulls}
\end{equation}
where scalars $X^i$ are the target space coordinates for the transverse 
space of the fundamental string.  

Note, the above action (\ref{f1probact}) describes the probe fundamental 
string moving in the background of fields produced by the source brane 
configuration.  Namely, $G_{MN}$ and $B_{MN}$ in Eqs. (\ref{f1probact}) 
and (\ref{pullbacks}) are the fields in Eq. (\ref{f1dpsgsol}) produced by 
the source.  Also, in the supergravity solution (\ref{f1dpsgsol}), the 
coordinates $\vec{x}$, $y$ and $\vec{z}$ correspond to the target space 
coordinates $\vec{X}$ of the probe fundamental string and are assumed to 
be functions of time $\tau=X^0$ only, i.e. $\vec{x}=\vec{x}(\tau)$.  
Namely, the probe fundamental string moves in the source background without 
oscillating.  Therefore, the probe action (\ref{f1probact}) takes the 
following form:
\begin{eqnarray}
S&=&-T_f\int d\tau d\sigma \left[\sqrt{-(-H^{-1}_FH^{-1/2}_p+
H^{-1/2}_pv^2_{\parallel}+H^{1/2}_pv^2_{\perp})H^{-1}_FH^{1/2}_p}
-H^{-1}_F\right]
\cr
&=&-T_f\int d\tau d\sigma\, H^{-1}_F\left[\sqrt{1-H_Fv^2_{\parallel}
-H_pH_Fv^2_{\perp}}-1\right]
\cr
&=&-m_f\int d\tau\, H^{-1}_F\left[\sqrt{1-H_Fv^2_{\parallel}
-H_pH_Fv^2_{\perp}}-1\right],
\label{f1dpbckprob}
\end{eqnarray}
where $v_{\parallel}$ and $v_{\perp}$ are respectively the speeds of the 
probe fundamental string in the longitudinal direction $\vec{x}$ and the 
transverse direction $\vec{z}$ of the D$p$-brane:
\begin{equation}
v_{\parallel}\equiv\sqrt{\sum^p_{i=1}({{dx_i}\over{d\tau}})^2},\ \ \ \ 
v_{\perp}\equiv\sqrt{\sum^{8-p}_{k=1}({{dz_k}\over{d\tau}})^2}.
\label{velocity}
\end{equation}
Since the configuration under consideration is assumed to be independent 
of the longitudinal coordinate $X^1=\sigma$ of the fundamental string, the 
integration (with the possible regularization) with respect to $\sigma$ in 
Eq. (\ref{f1dpbckprob}) just gives the volume factor, which combines with 
the tension $T_f$ of the probe fundamental string to give the mass $m_f$ 
of the probe fundamental string in the third line in Eq. (\ref{f1dpbckprob}).  
So, the above probe action $S$ effectively describes the dynamics of a test 
particle with mass $m_f$ moving in the background fields.  

In the core region of the fundamental string and the D$p$-brane ($x\approx 
0$ and $z\approx 0$) or in the large source charge limit ($Q_F\gg 1$ and 
$Q_p\gg 1$), the harmonic functions in the action (\ref{f1dpbckprob}) have 
the following forms:
\begin{equation}
H_F=\left\{\matrix{{{Q_F}\over{(x^2+4Q_pz)^3}},\ \ \ p<5 \cr 
{{Q_F}\over{(x^2+4Q_5z)^{7\over 2}}},\ \ \ p=5}\right.,\ \ \ \ 
H_p={{Q_p}\over z}.
\label{nearharms}
\end{equation}
In the case of the delocalized intersecting source configuration, the harmonic 
functions have the following forms:
\begin{equation}
H_F=1+{{Q_F}\over {z^{6-p}}},\ \ \ \ 
H_p=1+{{Q_p}\over {z^{6-p}}},
\label{delocalharms}
\end{equation}
where the constant terms 1 in the harmonic functions are absent in the 
near horizon region ($z\approx 0$).   Note, in the partially localized 
case with the harmonic functions given by Eq.(\ref{nearharms}), unlike the 
delocalized case with the harmonic functions (\ref{delocalharms}), the 
background geometry has the explicit dependence on the radial coordinate 
$x$ of the worldvolume space of the D$p$-brane.  This makes the study of 
non-trivial dynamics of the probe fundamental string in the relative 
transverse space possible.  

So, explicitly in terms of the parameters of the source supergravity 
solution, the probe action (\ref{f1dpbckprob}) in the core region of 
the constituent source branes takes the following form:
\begin{equation}
S=-m_f\int d\tau {{(x^2+4Q_pz)^{n\over 2}}\over{Q_F}}
\left[\sqrt{1-{{Q_Fv^2_{\parallel}}\over{(x^2+4Q_pz)^{n\over 2}}}
-{{Q_pQ_Fv^2_{\perp}}\over{z(x^2+4Q_pz)^{n\over 2}}}}-1\right], 
\label{prbact1}
\end{equation}
where $n=6$ [$n=7$] for $p<5$ [$p=5$], and 
\begin{equation}
S=-m_f\int d\tau {{z^{6-p}}\over{Q_F}}\left[\sqrt{1-
{{Q_Fv^2_{\parallel}}\over{z^{6-p}}}-{{Q_pQ_Fv^2_{\perp}}\over{z^{12-2p}}}}
-1\right],
\label{prbact3}
\end{equation}  
for the delocalized case.  

These actions for the probe fundamental string effectively describe 
the dynamics of a particle with mass $m_f$ moving in a velocity dependent 
potential.  When the motion of the probe is restricted either to the relative 
transverse space (with the coordinates $\vec{x}$) or to the overall 
transverse space (with the coordinates $\vec{z}$) of the source brane 
configuration, the force on the particle becomes central, as well.  

For the purpose of analyzing this motion, we set all the angular momenta of 
the probe except one in each space ($J_{\parallel}\neq 0$ in the relative 
transverse space and $J_{\perp}\neq 0$ in the overall transverse space) equal 
to zero.  And one introduces the polar coordinates $(x,\theta_{\parallel})$ 
and $(z,\theta_{\perp})$ in the rotation planes respectively associated with 
the angular momenta $J_{\parallel}$ and $J_{\perp}$.  Then, the velocities 
$v_{\parallel}$ and $v_{\perp}$ in the relative transverse space and the 
overall transverse space, defined in Eq. (\ref{velocity}), are decomposed as
\begin{equation}
v^2_{\parallel}=\dot{x}^2+x^2\dot{\theta}^2_{\parallel},\ \ \ \ 
v^2_{\perp}=\dot{z}^2+z^2\dot{\theta}^2_{\perp},
\label{veldecomp}
\end{equation}
where the dot denotes the differentiation with respect to the time 
coordinate $\tau$. 

Then, in general, the angular momenta $J_{\parallel}$ and $J_{\perp}$ and 
the energy $E$ of the probe are given by
\begin{eqnarray}
J_{\parallel}&=&p_{\theta_{\parallel}}={{\partial L}\over {\partial 
\dot{\theta_{\parallel}}}}={{m_fx^2\dot{\theta}_{\parallel}}\over
\sqrt{1-H_Fv^2_{\parallel}-H_pH_Fv^2_{\perp}}},
\cr
J_{\perp}&=&p_{\theta_{\perp}}={{\partial L}\over {\partial 
\dot{\theta_{\perp}}}}={{m_fH_pz^2\dot{\theta}_{\perp}}\over
\sqrt{1-H_Fv^2_{\parallel}-H_pH_Fv^2_{\perp}}},
\cr
E&=&H={{\partial L}\over{\partial v_{\parallel}}}v_{\parallel}+
{{\partial L}\over{\partial v_{\perp}}}v_{\perp}-L=
{{m_f}\over{H_F}}\left[{1\over{\sqrt{1-H_Fv^2_{\parallel}-H_pH_F
v^2_{\perp}}}}-1\right].
\label{angenerg}
\end{eqnarray}

So, explicitly the expressions for the angular momenta and the energy 
of the probe fundamental string in each case are as follows:
\begin{eqnarray}
J_{\parallel}&=&{{m_fx^2\dot{\theta}_{\parallel}}\over
{\sqrt{1-{{Q_Fv^2_{\parallel}}\over{(x^2+4Q_pz)^{n\over 2}}}
-{{Q_pQ_Fv^2_{\perp}}\over{z(x^2+4Q_pz)^{n\over 2}}}}}},
\cr
J_{\perp}&=&{{m_fQ_pz\dot{\theta}_{\perp}}\over
{\sqrt{1-{{Q_Fv^2_{\parallel}}\over{(x^2+4Q_pz)^{n\over 2}}}
-{{Q_pQ_Fv^2_{\perp}}\over{z(x^2+4Q_pz)^{n\over 2}}}}}},
\cr
E&=&{{m_f(x^2+4Q_pz)^{n\over 2}}\over{Q_F}}\left[{1\over
\sqrt{1-{{Q_Fv^2_{\parallel}}\over{(x^2+4Q_pz)^{n\over 2}}}
-{{Q_pQ_Fv^2_{\perp}}\over{z(x^2+4Q_pz)^{n\over 2}}}}}-1\right],
\label{angenerg1}
\end{eqnarray}
where $n=6$ [$n=7$] for $p<5$ [$p=5$], and 
\begin{eqnarray}
J_{\parallel}&=&{{m_fx^2\dot{\theta}_{\parallel}}\over
\sqrt{1-{{Q_Fv^2_{\parallel}}\over{z^{6-p}}}-{{Q_pQ_Fv^2_{\perp}}\over
{z^{12-2p}}}}},
\cr
J_{\perp}&=&{{m_fQ_pz^{p-4}\dot{\theta}_{\perp}}\over
\sqrt{1-{{Q_Fv^2_{\parallel}}\over{z^{6-p}}}-{{Q_pQ_Fv^2_{\perp}}\over
{z^{12-2p}}}}},
\cr
E&=&{{m_fz^{6-p}}\over{Q_F}}\left[{1\over\sqrt{1-{{Q_Fv^2_{\parallel}}
\over{z^{6-p}}}-{{Q_pQ_Fv^2_{\perp}}\over{z^{12-2p}}}}}-1\right],
\label{angenerg3}
\end{eqnarray}
for the delocalized case.

\subsection{The motion of the probe fundamental string in the relative 
transverse space}

In this subsection, we study the dynamics of the probe fundamental string 
whose motion is restricted to the relative transverse space of the source 
brane configuration.  In this case, $v_{\perp}=0$ and the coordinate $z$ 
is constant in time.  Generally, the angular momentum $J_{\parallel}$  in 
the relative transverse space and the energy $E$ of the probe fundamental 
string have the following forms:
\begin{equation}
J_{\parallel}={{m_fx^2\dot{\theta}_{\parallel}}\over
\sqrt{1-H_Fv^2_{\parallel}}},\ \ \ \ \ 
E={{m_f}\over{H_F}}\left[{1\over{\sqrt{1-H_Fv^2_{\parallel}}}}-1\right].
\label{longangenerg}
\end{equation}
So, explicitly the angular momentum and the energy for each case are as
follows:
\begin{eqnarray}
J_{\parallel}&=&{{m_fx^2\dot{\theta}_{\parallel}}\over
{\sqrt{1-{{Q_Fv^2_{\parallel}}\over{(x^2+4Q_pz)^{n\over 2}}}}}},
\cr
E&=&{{m_f(x^2+4Q_pz)^{n\over 2}}\over{Q_F}}\left[{1\over\sqrt{1-
{{Q_Fv^2_{\parallel}}\over{(x^2+4Q_pz)^{n\over 2}}}}}-1\right],
\label{angharmrt1}
\end{eqnarray}
where $n=6$ [$n=7$] for $p<5$ [$p=5$], and
\begin{eqnarray}
J_{\parallel}&=&{{m_fx^2\dot{\theta}_{\parallel}}\over
\sqrt{1-{{Q_Fv^2_{\parallel}}\over{z^{6-p}}}}},
\cr
E&=&{{m_fz^{6-p}}\over{Q_F}}\left[{1\over\sqrt{1-{{Q_Fv^2_{\parallel}}
\over{z^{6-p}}}}}-1\right],
\label{angharmrt3}
\end{eqnarray}  
for the delocalized case.  

From the above expression for the energy $E$ of the probe fundamental 
string, one obtains the following kinetic relation:
\begin{equation}
E={1\over 2}m_fv^2_{\parallel}+W(x);\ \ \ \ 
W(x)=E[1-{{1+{E\over{2m_f}}H_F}\over{(1+{E\over{m_f}}H_F)^2}}],
\label{prbkinrel}
\end{equation}
where the harmonic function $H_F$ is given by Eq. (\ref{nearharms}) or 
Eq. (\ref{delocalharms}).  By further using the expression for the angular 
momentum $J_{\parallel}$ in Eq. (\ref{longangenerg}), one obtains the 
following kinetic relation for the radial motion of the probe: 
\begin{equation}
E={1\over 2}m_f\dot{x}^2+V(x);\ \ \ \ 
V(x)=E[1-{{1+{E\over{2m_f}}H_F}\over{(1+{E\over{m_f}}H_F)^2}}]+
{{J^2_{\parallel}}\over{2m_fx^2}}{1\over{(1+{E\over {m_f}}H_F)^2}}.
\label{prbradkinrel}
\end{equation}
So, the radial motion (along the $x$ direction) of the probe is that of the 
test particle with mass $m_f$ moving in a velocity-independent central 
force potential $V(x)$.  

In the partially localized case, the effective potential $V(x)$ 
is explicitly given by
\begin{equation}
V(x)=E[1-{{1+{{b^n}\over{2(x^2+4Q_pz)^{n\over 2}}}}\over{(1+{{b^n}\over
{(x^2+4Q_pz)^{n\over 2}}})^2}}]+{{J^2_{\parallel}}\over{2m_fx^2}}
{1\over{(1+{{b^n}\over{(x^2+4Q_pz)^{n\over 2}}})^2}},
\label{effpotreltr}
\end{equation}
where $b$ is the characteristic scale given by
\begin{equation}
b=\left({{EQ_F}\over{m_f}}\right)^{1\over n},
\label{chscl}
\end{equation}
where $n=6$ [$n=7$] for $p<5$ [$p=5$].  

The dynamics of the probe fundamental string along the radial direction $x$ 
can be studied by analyzing an effective velocity-dependent central force 
potential $V(x)$ in Eq. (\ref{prbradkinrel}).  In the ``delocalized'' 
source background with the harmonic function $H_F$ in Eq. 
(\ref{delocalharms}) being independent of the radial coordinate $x$, the 
dynamics of the probe fundamental string in the worldvolume space of the 
D$p$-brane is trivial: the only force on the probe is the repulsive 
centrifugal force due to non-zero angular momentum $J_{\parallel}$ of the 
probe.  However, with the partially localized source background, one can 
study non-trivial dynamics of the probe fundamental string since the source 
fundamental string is now localized on the source D$p$-brane and therefore 
the effective potential $V(x)$ in Eq. (\ref{effpotreltr}) has explicit 
dependence on $x$.  

At large distance $x\gg b$ from the source fundamental string, the effective 
potential in Eq. (\ref{effpotreltr}) takes the following form:
\begin{equation}
V_{x\gg b}\to {{3E^2Q_F}\over{2m_f(x^2+4Q_pz)^{n\over 2}}}+
{{J^2_{\parallel}}\over{2m_fx^2}}.
\label{lgdstptrelt}
\end{equation}
So, the motion of the probe fundamental string is qualitatively similar 
to the motion in the background of the source fundamental string only, except 
that the strength of the repulsive potential (the first term in Eq. 
(\ref{lgdstptrelt})) is decreased due to the presence of the source 
D$p$-brane.  This contribution from the source D$p$-brane gets enhanced for 
larger D$p$-brane charge $Q_p$ and at larger distance $z$ from the source 
D$p$-brane.  However, the centrifugal potential (the second term) remains 
the same regardless of the presence of the source D$p$-brane.  

At short distance $x\ll b$ from the source fundamental string and very close 
to the source D$p$-brane ($z\ll b^2/Q_p$), the effective potential is 
approximated to
\begin{equation}
V_{x\ll b,\,z\ll b^2/Q_p}\to E-{{m_f}\over{2Q_F}}(x^2+4Q_pz)^{n\over 2}+
{{m_fJ^2_{\parallel}}\over{2E^2Q^2_F}}{{(x^2+4Q_pz)^n}\over{x^2}}.
\label{stdstptrelt}
\end{equation}
The (energy $E$ independent) repulsive potential term is enhanced again 
due to the presence of the source D$p$-brane: for larger $Q_p$ and $z$, 
the repulsive force becomes stronger.  Unlike the case of the long distance 
region, the probe fundamental string now feels the effect of the source 
D$p$-brane on the centrifugal potential when the probe fundamental string 
gets very close to the source.  

We now discuss the motion of the probe fundamental string in the source 
background.  The turning points, where the radial velocity $\dot{x}$ of 
the probe becomes zero, are located at the values of $x$ where $E=V(x)$ 
(Cf. Eq. (\ref{prbradkinrel})) and therefore are the roots of the following 
equation:
\begin{equation}
1+{{b^n}\over{2(x^2+4Q_pz)^{n\over 2}}}={{b^2_*}\over{x^2}},
\label{reltrntp}
\end{equation}
where $b$ is given in Eq. (\ref{chscl}), $b_*\equiv J_{\parallel}/
\sqrt{2m_fE}$ and again $n=6$ [$n=7$] for $p<5$ [$p=5$].  
This equation always has one positive root $x$ for non-zero angular momentum 
$J_{\parallel}$ of the probe.  And when $J_{\parallel}=0$, there is no 
root $x$ for this equation.  Furthermore, the radial force $F(x)=-{{dV(x)}
\over{dx}}$ on the probe diverges as $x\to 0$, when $J_{\parallel}\neq 0$.  
On the other hand, when the angular momentum $J_{\parallel}$ is zero, the 
force vanishes at $x=0$.  So, the motion of the probe fundamental string 
can be summarised as follows.  When the probe has non-zero angular momentum, 
the probe will always be scattered away when it reaches the source fundamental 
string.  When the probe has no angular momentum, it will eventually be 
absorbed by the source.  

In the following, we compare the dynamics of the BIons, which is 
previously studied in Refs. \cite{gp1,blm,gp2}, to the dynamics 
of the probe fundamental string along the worldvolume direction of 
the D-brane studied in the above paragraphs.  It is natural to expect 
that these two systems have the same dynamics, since the BIons in 
the $(p+1)$-dimensional DBI theory are interpreted as the ends of fundamental 
strings on D$p$-brane in the very weak string coupling limit ($g_s\to 0$).  
Namely, the motion of the ends of the fundamental strings on the D-brane 
along the D-brane worldvolume direction is essentially the motion of 
the BIons.  However, as we shall see in the following, our description of 
dynamics of the probe fundamental string studied in the above is too 
simplified to reproduce the dynamics of the worldvolume solitons, i.e. BIons.  

In the following, we summarize the dynamics of the BIons, studied in Refs. 
\cite{gp1,blm,gp2}, for the purpose of comparing the dynamics of BIons to 
the dynamics of the probe and for the purpose of fixing the notations.  
The $(p+1)$-dimensional DBI action has the following form:
\begin{equation}
S_{\rm DBI}=-\int d^{p+1}\sigma \sqrt{-\det(\eta_{MN}\partial_{\mu}X^M
\partial_{\nu}X^N+F_{\mu\nu})},
\label{biact}
\end{equation}
where $\eta_{MN}$ ($M,N=0,1,...,9$) is the metric for the Minkowskian 
target space and $F_{\mu\nu}=\partial_{\mu}A_{\nu}-\partial_{\nu}A_{\mu}$ 
($\mu,\nu=0,1,...,p$) is the field strength of the worldvolume $U(1)$ 
gauge field $A_{\mu}$.  In the `static' gauge, in which the worldvolume 
coordinates $\sigma^{\mu}$ are identified with the target space coordinates 
as $\sigma^{\mu}=X^{\mu}$ ($(X^M)=(X^{\mu},X^m)$ with $m=p+1,...,9$), the 
DBI action (\ref{biact}) takes the following form:
\begin{equation}
S_{\rm DBI}=-\int d^{p+1}\sigma \sqrt{-\det(\eta_{\mu\nu}+\partial_{\mu}X^m
\partial_{\nu}X^m+F_{\mu\nu})}.
\label{statbiact}
\end{equation}

The ends of the fundamental strings on the D$p$-brane worldvolume 
correspond to the BIons which carry the electric charge of the worldvolume 
$U(1)$ gauge field $A_{\mu}$.  One scalar, say $X:=X^{p+1}$, associated 
with the longitudinal direction of the fundamental string in the target 
space is non-trivial.  In general, the BIon solution has the following form 
\cite{cm,gib}:
\begin{equation}
A_0=-H,\ \ \ X=H;\ \ \ \ 
H=\sum_k{{q_k}\over{|\vec{\sigma}-\vec{x}_k|^{p-2}}},
\label{bionsol}
\end{equation}
where $\vec{\sigma}=(\sigma^1,...,\sigma^p)$ is the spatial components of 
the worldvolume coordinates $(\sigma^{\mu})=(\tau,\vec{\sigma})$ 
and $\vec{x}_k$ is the location of a BIon with the electric charge $q_k$ 
in the worldvolume space.  

Since we are interested in the low velocity dynamics of BIons, it is 
sufficient to consider the following linearized approximation to the 
DBI action (\ref{statbiact}):
\begin{eqnarray}
S_{\rm DBI}&\approx&{1\over 2}\int d^{p+1}\sigma[\delta_{mn}\eta^{\mu\nu}
\partial_{\mu}X^m\partial_{\nu}X^n+{1\over 2}F_{\mu\nu}F^{\mu\nu}]
\cr
&=&{1\over 2}\int d^{p+1}\sigma[\eta^{\mu\nu}\partial_{\mu}X\partial_{\nu}X
+{1\over 2}F_{\mu\nu}F^{\mu\nu}],
\label{lindbiact}
\end{eqnarray}
where in the second line only one scalar $X$ associated with the longitudinal 
direction of the attached fundamental string is kept.  

To study the dynamics of the BIons, we allow the locations of the BIons to 
change with time, i.e. $\vec{x}_k=\vec{x}_k(\tau)$.  So, $\vec{v}_k=
{{d\vec{x}_k(\tau)}\over{d\tau}}$ is the velocity of the $k$-th BIon with 
the electric charge $q_k$.  
One has to also add the following source term $S_{\rm source}$ for the BI 
$U(1)$ field $A_{\mu}$ and scalar field $X$, and the free term $S_{\rm free}$ 
for the BIons:
\begin{eqnarray}
S_{\rm source}&=&(2-p)\Omega_{p-1}\sum_k\int d\tau\,[q_kX\sqrt{1-v^2_k}
+q_kA_{\mu}
{{\partial x^{\mu}_k}\over{\partial \tau}}],
\cr
S_{\rm free}&=&-\sum_k\int d\tau m_k(\epsilon)\sqrt{1-v^2_k},
\label{sourceact}
\end{eqnarray}
where $\Omega_{p-1}=2\pi^{p\over 2}/\Gamma({p\over 2})$ is the volume 
of the unit ($p-1$)-sphere $S^{p-1}$ and $m_k(\epsilon)$ is the `regularized' 
mass of the $k$-th BIon.  Here, $\epsilon$ is the cut-off for the 
radial distance from the BIons, i.e. we restrict ourselves to the region 
$|\vec{\sigma}-\vec{x}_k|\geq\epsilon$.  Then, $m_k(\epsilon)$ corresponds 
to the mass of the fundamental string whose length is truncated due to 
the regularization \cite{cm}.  The source term can be interpreted as being 
related to the bulk supergravity configuration.  Namely, the D$p$-brane, whose 
shape in the $(x,X)$-plane is given by $X(x)$, is the source of the first 
term in $S_{\rm source}$, and the the end of the fundamental string on the 
D$p$-brane is the source of the worldvolume $U(1)$ gauge field $A_{\mu}$ 
(the second term in $S_{\rm source}$ describes such coupling).  $S_{\rm free}$ 
is the action for the BIons with masses $m_k(\epsilon)$.  

Note, since the BIons, which carry electric charges, have non-zero velocities, 
the magnetic field is induced and the velocity dependent force will 
also be induced.  To study such and other effects on the BIon dynamics due to 
non-zero velocities, one perturbs the fields around the static configuration 
(\ref{bionsol}).  Then, one substitutes the perturbed fields, which 
satisfy the equations of motion to the order ${\cal O}(v^2)$ in the 
velocity $v$ of the BIons, into the  action $S=S_{\rm DBI}+S_{\rm source}+
S_{\rm free}$ in order to obtain the following on-shell effective action 
for the BIons to the order ${\cal O}(v^2)$ \cite{gp1,blm,gp2}:
\begin{equation}
S\approx {{p-2}\over 2}\Omega_{p-1}\int d\tau [\sum_km_k(\epsilon)v^2_k+
(p-2)\Omega_{p-1}\sum_{k<\ell}q_kq_{\ell}{{|\vec{v}_k-\vec{v}_{\ell}|^2}
\over{|\vec{x}_k-\vec{x}_{\ell}|^{p-2}}}],
\label{onshellbi}
\end{equation}
for large separations $|\vec{x}_k-\vec{x}_{\ell}|\gg 0$ for BIons.  

The system of the probe fundamental string (with the action (\ref{f1probact})) 
moving in the source background, given by Eq. (\ref{f1dpsgsol}) with Eq. 
(\ref{coinharms}), is the bulk counterpart to the dynamic system 
of two BIons in the $(p+1)$-dimensional DBI theory.
In this case, the indices $k$ and $\ell$ in Eqs. (\ref{bionsol}) and 
(\ref{onshellbi}) run from 1 to 2.  We let the first [second] BIon  
correspond to the probe [the source] fundamental string.  
This is a two-body system under a central force.  Such system can be reduced 
to an equivalent one-body system by replacing the positions $\vec{x}_1$ and 
$\vec{x}_2$ of the BIons by their center-of-mass position $\vec{R}=
(m_1\vec{x}_1+m_2\vec{x}_2)/(m_1+m_2)$ and relative position $\vec{r}=
\vec{x}_1-\vec{x}_2$.  The action (\ref{onshellbi}) then transforms to the 
following form:
\begin{equation}
S\approx {{p-2}\over 2}\Omega_{p-1}\int d\tau [MV^2+\mu v^2+(p-2)\Omega_{p-1}
{{q_1q_2v^2}\over{r^{p-2}}}],
\label{2bdyact}
\end{equation}
where $M=m_1+m_2$ is the total mass, $\mu=m_1m_2/M$ is the reduced mass, 
$V=|{{d\vec{R}}\over{d\tau}}|$ is the center-of-mass velocity, and $v=
|{{d\vec{r}}\over{d\tau}}|$ is the relative velocity.  Thus, the 
motion of BIons is described by the geodesic motion in the moduli space 
with the following metric:
\begin{equation}
ds^2_{MS}=Md\vec{R}^2+[\mu+(p-2)\Omega_{p-1}{{q_1q_2}\over{r^{p-2}}}]
d\vec{r}^2.
\label{modmetrc}
\end{equation}
As expected, the center-of-mass moves freely but the relative motion of 
the BIons is under the influence of a velocity dependent central potential.  
Since the source fundamental string is assumed to be much heavier than the 
probe fundamental string, the second BIon is much heavier than the first 
BIon, i.e. $m_1\ll m_2$.  Then, the action (\ref{2bdyact}) in the 
center-of-mass coordinate system ($\vec{V}=\vec{0}$) is approximated to
\footnote{When $m_1\ll m_2$, the quantities in the center of mass 
frame are approximated as $\vec{x}_1={{m_2}\over{m_1+m_2}}\vec{r}
\approx\vec{r}$ and $\mu={{m_1m_2}\over{m_1+m_2}}\approx m_1$.}:
\begin{equation}
S\approx {{p-2}\over 2}\Omega_{p-1}\int d\tau [m_1v^2_1+(p-2)\Omega_{p-1}
{{q_1q_2v^2_1}\over{r^{p-2}}}],
\label{aprx2bdyact}
\end{equation}
which describes dynamics of the first BIon in the background of the second 
BIon, which is fixed in space.  

In order to compare the above result for the BIon dynamics to the bulk theory 
result, one has to obtain the velocity dependent potential from the energy 
$E$ of the probe fundamental string in Eq. (\ref{angharmrt1}).  
In the limit of very small probe velocity ($v_{\parallel}\approx 0$), 
the energy $E$ of the probe fundamental string in Eq. (\ref{angharmrt1}) is 
expanded in powers of the velocity as
\begin{equation}
E={1\over 2}m_fv^2_{\parallel}+{{3m_fQ_Fv^4_{\parallel}}\over
{8(x^2+4Q_pz)^{n\over 2}}}+{{5m_fQ^2_Fv^6_{\parallel}}\over
{16(x^2+4Q_pz)^n}}+...,
\label{relenergexp}
\end{equation}
and, therefore, to the leading order in $v_{\parallel}$, the velocity 
dependent potential on the probe is
\begin{equation}
V_{\rm eff}\approx {{3m_fQ_Fv^4_{\parallel}}\over{8(x^2+4Q_pz)^{n\over 2}}}.
\label{veleffpot}
\end{equation}

This expression for the effective potential on the probe fundamental 
string has different dependence on the velocity $v_{\parallel}$ and the 
radial coordinate $x$ from the effective potential on the light BIon 
given in Eq. (\ref{aprx2bdyact}).  In addition, one also finds disagreement 
of the probe moduli metric, which describes the geodesic motion of the probe 
fundamental string, with the moduli metric (\ref{modmetrc}) of BIons, as 
we see in the following.  In the limit of very small probe velocity 
($v_{\parallel}\approx 0$ and $v_{\perp}\approx 0$), the on-shell action 
(\ref{f1dpbckprob}) is approximated to
\begin{equation}
S\approx {1\over 2}m_f\int d\tau\,(v^2_{\parallel}+H_pv^2_{\perp}), 
\label{smallvelact}
\end{equation}
to the lowest order in the velocities.  The vanishing of the static potential 
in this on-shell action is in accordance with the fact that we are 
considering a BPS configuration.  From the definitions of the probe 
fundamental string velocities (\ref{velocity}), one can see that the moduli 
metric of the probe fundamental string is given by
\begin{equation}
ds^2_F=dx_idx_i+H_pdz_kdz_k.
\label{f1modmet}
\end{equation}
This moduli metric implies that the probe moves freely along the 
worldvolume directions of the D$p$-brane, whereas the moduli metric 
(\ref{modmetrc}) for the BIons describes the motion under the 
influence of the velocity dependent central potential.  

This disagreement may be traced from the following factors.  
First, in calculating the effective action for the probe fundamental string, 
we assumed that the field configurations are static, uninfluenced by the 
moving probe fundamental string.  As was done originally in Refs. 
\cite{man,ah1,ah2,gr,fe}, when one studies the motion of collection of 
interacting solitons (in the low-velocity limit), which is described by 
the geodesic motion in the moduli space, one usually takes into account 
the perturbation (in a slow-motion expansion) of the original static fields 
due to non-zero velocities of the solitons.  This is properly done in the 
case of the interaction of BIons in the above, but not in the case of the 
probe fundamental string moving in the source background.  Namely, in the 
source-probe method, one assumes that source is much heavier than the probe 
and therefore the source is uninfluenced by the probe.  On the other hand, 
in the case of BIon dynamics, first we assumed that all the BIons have 
the comparable mass (therefore, the field produced by one BIon is influenced 
by those of other BIons) and at the end we let one of the BIons to be much 
heavier and the others.  Second, the supergravity background (\ref{f1dpsgsol}) 
in which the probe fundamental string moves corresponds to the 
configuration where the source D$p$-brane is delocalized along the 
longitudinal direction (the $y$ direction) of the source fundamental 
string.  On the other hand, as for the BIon solution in Eq. 
(\ref{bionsol}), the location of the D$p$-brane along the 
longitudinal direction of the fundamental string is specified by the 
scalar $X$.  So, although the partially localized supergravity solution 
(\ref{f1dpsgsol}) has all the parameters of the BIons, i.e. the charges 
and the locations of the BIons, it still lacks one special feature of 
the BIon solution that a scalar $X$ of the BIon solution specifies the 
location or the shape of the D$p$-brane along the longitudinal direction 
of the fundamental string.  In the fully localized supergravity solution, 
one would expect to see the shape of D$p$-brane pulled by the fundamental 
string, just like the case of the BIon solution.  In the case of the 
configurations describing one type of brane within the worldvolume of 
another type of brane, we will not encounter with this problem, since 
there is no relative transverse direction that is delocalized.  If one 
properly takes into account the above observations, it might be possible 
to reproduce the moduli metric for the BIon interaction by studying probe 
fundamental string moving in the source background of the fundamental 
strings ending on D$p$-brane.

\subsection{The motion of the probe fundamental string in the overall 
transverse space}

In this subsection, we study the dynamics of the probe fundamental string 
whose motion is restricted to the overall transverse space.  In this case, 
$v_{\parallel}=0$ and the coordinate $x$ is constant in time.  Generally, 
the angular momentum $J_{\perp}$  in the overall transverse space and 
the energy $E$ of the probe fundamental string have the following forms:
\begin{equation}
J_{\perp}={{m_fH_pz^2\dot{\theta}_{\perp}}\over\sqrt{1-H_pH_Fv^2_{\perp}}},
\ \ \ \ \ 
E={{m_f}\over{H_F}}\left[{1\over{\sqrt{1-H_pH_Fv^2_{\perp}}}}-1\right].
\label{tranangenerg}
\end{equation}
So, the explicit expressions for the angular momentum and the energy are
\begin{eqnarray}
J_{\perp}&=&{{m_fQ_pz\dot{\theta}_{\perp}}
\over{\sqrt{1-{{Q_pQ_Fv^2_{\perp}}\over{z(x^2+4Q_pz)^{n\over 2}}}}}},
\cr
E&=&{{m_f(x^2+4Q_pz)^{n\over 2}}\over{Q_F}}\left[{1\over
\sqrt{1-{{Q_pQ_Fv^2_{\perp}}\over{z(x^2+4Q_pz)^{n\over 2}}}}}-1\right],
\label{angharmot1}
\end{eqnarray}
where $n=6$ [$n=7$] for $p<5$ [$p=5$], and
\begin{eqnarray}
J_{\perp}&=&{{m_fQ_pz^{p-4}\dot{\theta}_{\perp}}\over
\sqrt{1-{{Q_pQ_Fv^2_{\perp}}\over{z^{12-2p}}}}}
\cr
E&=&{{m_fz^{6-p}}\over{Q_F}}\left[{1\over\sqrt{1-{{Q_pQ_Fv^2_{\perp}}\over
{z^{12-2p}}}}}-1\right],
\label{angharmot3}
\end{eqnarray}
for the delocalized case.  

From the above expression for the energy $E$ of the probe fundamental 
string, one obtains the following kinetic relation:
\begin{equation}
E={1\over 2}m_fv^2_{\perp}+W(z);\ \ \ \ 
W(z)=E[1-{1\over{H_p}}{{1+{E\over{2m_f}}H_F}\over{(1+{E\over {m_f}}
H_F)^2}}],
\label{prbkinreltr}
\end{equation}
where the harmonic functions $H_F$ and $H_p$ are given in Eqs. 
(\ref{nearharms}) and (\ref{delocalharms}).  
Further using the expression for the angular momentum $J_{\perp}$ in 
Eq. (\ref{tranangenerg}), one obtains the following kinetic relation 
for the radial motion of the probe:
\begin{equation}
E={1\over 2}m_f\dot{z}^2+V(z);\ \ \ \ 
V(z)=E[1-{1\over{H_p}}{{1+{E\over{2m_f}}H_F}\over{(1+{E\over{m_f}}H_F)^2}}]+
{{J^2_{\perp}}\over{2m_fH^2_pz^2}}{1\over{(1+{E\over{m_f}}H_F)^2}}.
\label{prbradkinreltr}
\end{equation}
So, again the radial motion (along the $z$ direction) of the probe 
fundamental string is that of a test particle with mass $m_f$ moving in 
an effective velocity-independent central force potential $V(z)$.  

We notice the qualitative difference in the effective potential 
(therefore, the qualitative difference in the dynamics of the probe) between  
this case and the case of the dynamics in the relative transverse space of 
the source brane configuration with the effective potential given in Eq. 
(\ref{prbradkinrel}).  Namely, whereas the effective potential for the 
dynamics in the worldvolume direction of the source D$p$-brane is affected 
by the source D$p$-brane only through the D$p$-brane charge $Q_p$ in the 
harmonic function $H_F$ for the source fundamental string, in the case 
of the dynamics in the directions transverse to the D$p$-brane the 
effective potential explicitly depends on the harmonic function $H_p$ of 
the D$p$-brane.  This is expected from the fact that our supergravity 
field background (\ref{f1dpsgsol}) for the source is delocalized along 
the longitudinal direction of the fundamental string: the probe 
fundamental string will feel the uniform force field of the same strength 
(produced by the source D$p$-brane) as it moves along the D$p$-brane 
worldvolume direction.  However, the source D$p$-brane is in fact pulled 
by the source fundamental string and therefore the probe 
fundamental string will feel the varying force field of the source 
D$p$-brane and ultimately hit the D$p$-brane as it moves towards 
the source fundamental string along the D$p$-brane worldvolume 
direction.  This force on the probe fundamental string due to the source 
D$p$-brane is the bulk counterpart to the force on the BIon due to the 
scalar charge of $X$ in Eq. (\ref{bionsol}).  
This is one of the reasons for the mismatch of the probe dynamics 
in the D$p$-brane worldvolume direction and the BIon dynamics, as pointed 
out in the previous section.  

For the partially localized case, the effective potential in the core 
region is explicitly given by
\begin{equation}
V(z)=[1-{z\over{Q_p}}{{1+{{(Q_pb)^{n\over 2}}\over{2(x^2+4Q_pz)^{n\over 
2}}}}\over{(1+{{(Q_pb)^{n\over 2}}\over{(x^2+4Q_pz)^{n\over 2}}})^2}}]
+{{J^2_{\perp}}\over{2m_fQ^2_p}}{1\over{(1+{{(Q_pb)^{n\over 2}}\over
{(x^2+4Q_pz)^{n\over 2}}}})^2},
\label{effpotovlpt}
\end{equation}
where the characteristic scale $b$ has the following form:
\begin{equation}
b=\left({{EQ_F}\over{m_fQ^{n\over 2}_p}}\right)^{2\over n},
\label{partchscal}
\end{equation}
and for the delocalized case, 
\begin{equation}
V(z)=E[1-{{z^{6-p}}\over{Q_p}}{{1+{{b^{6-p}}\over{2z^{6-p}}}}\over
{(1+{{b^{6-p}}\over{z^{6-p}}})^2}}]+{{J^2_{\perp}z^{10-2p}}\over
{2m_fQ^2_p}}{1\over{(1+{{b^{6-p}}\over{z^{6-p}}})^2}},
\label{effpotovldel}
\end{equation}
where the characteristic scale $b$ is given by:
\begin{equation}
b=\left({{EQ_F}\over{m_f}}\right)^{1\over{6-p}}.
\label{deltchscal}
\end{equation}

We now analyze the dynamics of the probe fundamental string moving along the 
radial direction $z$ of the overall transverse space.  The dynamics is 
non-trivial for both delocalized and partially localized cases.  We study 
both of these cases and compare the differences.  

At large distance $z\gg b$ from the source, the effective potential $V(z)$ 
in Eq. (\ref{prbradkinreltr}) takes the following form:
\begin{equation}
V_{z\gg b}\to E[1-{z\over{Q_p}}(1-{{3EQ_F}\over{2m_f(x^2+4Q_pz)^{n\over 2}}})]
+{{J^2_{\perp}}\over{2m_fQ^2_p}}[1-{{2EQ_F}\over{m_f(x^2+4Q_pz)^{n\over 2}}}],
\label{lgdstptovlt}
\end{equation}
for the partially localized case, and
\begin{equation}
V_{z\gg b}\to E[1-{{z^{6-p}}\over{Q_p}}(1-{{3EQ_F}\over{2m_fz^{6-p}}})]
+{{J^2_{\perp}z^{10-2p}}\over{2m_fQ^2_p}}(1-{{2EQ_F}\over{m_fz^{6-p}}}),
\label{delgdstptovlt}
\end{equation}
for the delocalized case.  Due to the presence of the D$p$-brane, the 
usual repulsive potential of the source fundamental string (the part of 
the potential which is independent of $J_{\perp}$) gets an additional 
potential contribution from the source D$p$-brane.  In the delocalized 
case (Eq. (\ref{delgdstptovlt})), the repulsive force due to the probe 
fundamental string is completely ``screened'' by the source D$p$-brane 
and the probe feels the repulsive force due to the D$p$-brane, only.  
However, when the source fundamental string is localized at the source 
D$p$-brane, the probe feels some repulsive contribution in the effective 
potential which signals existence of the source fundamental string.   
The (repulsive) centrifugal potential ($J_{\perp}$ dependent term) on 
the probe is again suppressed due to the presence of the D$p$-brane and 
is not repulsive any more.  (The term ${{J^2_{\perp}}\over{2m_fH^2_pz^2}}$ 
in Eq. (\ref{prbradkinreltr}) becomes a standard centrifugal potential 
term, if the source D$p$-brane is absent, i.e. $H_p=1$.)  In the 
case of the partially localized case, this term does not give rise to the 
force on the probe (ignoring the subleading $Q_F$ dependent term).  
However, when the source fundamental string is delocalized, the probe 
still feels a $J_{\perp}$ dependent attractive force (to the leading 
order, ignoring the subleading $Q_F$ dependent term).  

At short distance $z\ll b$ from the source D$p$-brane and very close to the 
source fundamental string ($x\ll\sqrt{Q_pb}$), the effective potential 
is approximated to
\begin{equation}
V_{z\ll b,\,x\ll\sqrt{Q_pb}}\to E-{{m_f}\over{2Q_pQ_F}}z(x^2+4Q_pz)^{n\over 2}
+{{m_fJ^2_{\perp}}\over{2E^2Q^2_pQ^2_F}}(x^2+4Q_pz)^n,
\label{shdstptovlt}
\end{equation}
for the partially localized case, and
\begin{equation}
V_{z\ll b}\to E-{{m_f}\over{2Q_pQ_F}}z^{12-2p}+{{m_fJ^2_{\perp}}\over
{2E^2Q^2_pQ^2_F}}z^{22-4p},
\label{delshdstptovlt}
\end{equation}
for the delocalized case.  
The usual ($J_{\perp}$ independent) repulsive potential term due to 
the source fundamental string is suppressed by the contribution from the 
source D$p$-brane.  Again, the repulsive centrifugal potential is 
completely suppressed and becomes attractive.  

We now study the motion of the probe along the radial direction $z$ of the 
overall transverse space.  The turning point $z$, where the radial velocity 
$\dot{z}$ vanishes, of the probe's motion is given by the root of the 
following equation:
\begin{equation}
1+{{(Q_pb)^{n\over 2}}\over{2(x^2+4Q_pz)^{n\over 2}}}=
{{b^2_*}\over{Q_p}}{1\over z},
\label{ovtrntp}
\end{equation}
where $b$ is given in Eq. (\ref{partchscal}) and $b_*=J_{\perp}/\sqrt{2m_fE}$. 
In the delocalized case, the turning point $z$ satisfies the following equation:
\begin{equation}
1+{{b^{6-p}}\over{2z^{6-p}}}={{b^2_*}\over{Q_p}}z^{4-p},
\label{delovtrntp}
\end{equation}
where $b$ is given in Eq. (\ref{deltchscal}).
In the partially localized case, there is always one turning point 
at positive $z$ when $J_{\perp}\neq 0$, whereas there is no turning point 
for the $J_{\perp}=0$ case.  In the delocalized case, when the angular 
momentum is non-zero, there is ($i$) one turning point at positive $z$ for 
$p<4$, ($ii$) one turning point at positive $z$ [no turning point] for a 
sufficiently large [small] value of $J_{\perp}$ for $p=4$, and ($iii$) one 
turning point at $z=b^2_*/Q_p-b/2$ ($b^2_*>Q_pb/2$) for $p=5$, whereas there 
is no turning point for the $J_{\perp}=0$ case.  Furthermore, the force on 
the probe along the radial direction $z$, i.e. $F(z)=-{{dV(z)}\over{dz}}$, 
is always positive [vanishes] at $z=0$ when $x>0$ [$x=0$] in the partially 
localized case.  In the delocalized case, the force always vanishes at $z=0$.  
So, the motion of the probe fundamental string along the $z$ direction can be  
summarized as follows.  In the partially localized case, away from the 
source fundamental string ($x>0$), the probe fundamental string will 
always bounce back as it approaches the source D$p$-brane, but can be 
eventually absorbed by the source D$p$-brane when the probe 
approaches the D$p$-brane inside of the worldvolume of the source 
fundamental string ($x=0$).  In the delocalized case, the probe with 
$J_{\perp}=0$ will always be absorbed by the source D$p$-brane.  
This seems to be due to the fact that the source fundamental string 
is delocalized on the source D$p$-brane, i.e. is uniformly distributed 
over the worldvolume of the D$p$-brane.  When the probe has non-zero 
angular momentum $J_{\perp}$, the probe will bounce back as it 
approaches the D$p$-brane for the following cases: ($i$) $p<4$, ($ii$) the 
sufficiently large value of $J_{\perp}$ with $p=4$, and ($iii$) 
$b^2_*>Q_pb/2$ with $p=5$.  Otherwise, the probe will always be absorbed 
by the source as it approaches the D$p$-brane.

\end{document}